\documentclass[12pt]{iopart}

\usepackage[utf8]{inputenc}
\expandafter\let\csname equation*\endcsname\relax
\expandafter\let\csname endequation*\endcsname\relax
\usepackage{amsmath}
\usepackage{amsfonts}
\usepackage{amssymb}
\usepackage{footnote}
\usepackage{caption}
\usepackage{fancyvrb}
\usepackage{makecell}
\usepackage{graphicx}
\usepackage{verbatim}

\usepackage[backend=bibtex,bibstyle=ieee,citestyle=numeric-comp]{biblatex}
\begin{document}

\title[PICSAR-QED]{PICSAR-QED: a Monte Carlo module to simulate Strong-Field Quantum Electrodynamics in Particle-In-Cell codes for exascale architectures}
\author{Luca Fedeli}
\address{
Lasers, Interaction and Dynamics Laboratory, CEA Saclay, Paris-Saclay University
}

\author{Neïl Zaïm}
\address{
Lasers, Interaction and Dynamics Laboratory, CEA Saclay, Paris-Saclay University
}

\author{Antonin Sainte-Marie}
\address{
Lasers, Interaction and Dynamics Laboratory, CEA Saclay, Paris-Saclay University
}

\author{Maxence Thévenet}
\address{
Deutsches Elektronen-Synchrotron , Notkestraße 85, D-22607 Hamburg
}

\author{Axel Huebl}
\address{
Lawrence Berkeley National Laboratory, Berkeley, CA 94720, USA
}

\author{Andrew Myers}
\address{
Lawrence Berkeley National Laboratory, Berkeley, CA 94720, USA
}

\author{Jean-Luc Vay}
\address{
Lawrence Berkeley National Laboratory, Berkeley, CA 94720, USA
}

\author{Henri Vincenti}
\address{
Lasers, Interaction and Dynamics Laboratory, CEA Saclay, Paris-Saclay University
}

\vspace{10pt}
\begin{indented}
\item[]\date{}
\end{indented}

\begin{abstract}
Physical scenarios where the electromagnetic fields are so strong that Quantum ElectroDynamics (QED) plays a substantial role are one of the frontiers of contemporary plasma physics research.
Investigating those scenarios requires state-of-the-art Particle-In-Cell (PIC) codes able to run on top high-performance computing machines and, at the same time, able to simulate strong-field QED processes.
This work presents the PICSAR-QED library, an open-source, portable implementation of a Monte Carlo module designed to provide modern PIC codes with the capability to simulate such processes, and optimized for high-performance computing. Detailed tests and benchmarks are carried out to validate the physical models in PICSAR-QED, to study how numerical parameters affect such models, and to demonstrate its capability to run on different architectures (CPUs and GPUs). 
Its integration with WarpX, a state-of-the-art PIC code designed to deliver scalable performance on upcoming exascale supercomputers, is also discussed and validated against results from the existing literature. 
\end{abstract}

\section{Introduction}

\begin{figure}
    \centering
    \includegraphics[width=0.8\columnwidth]{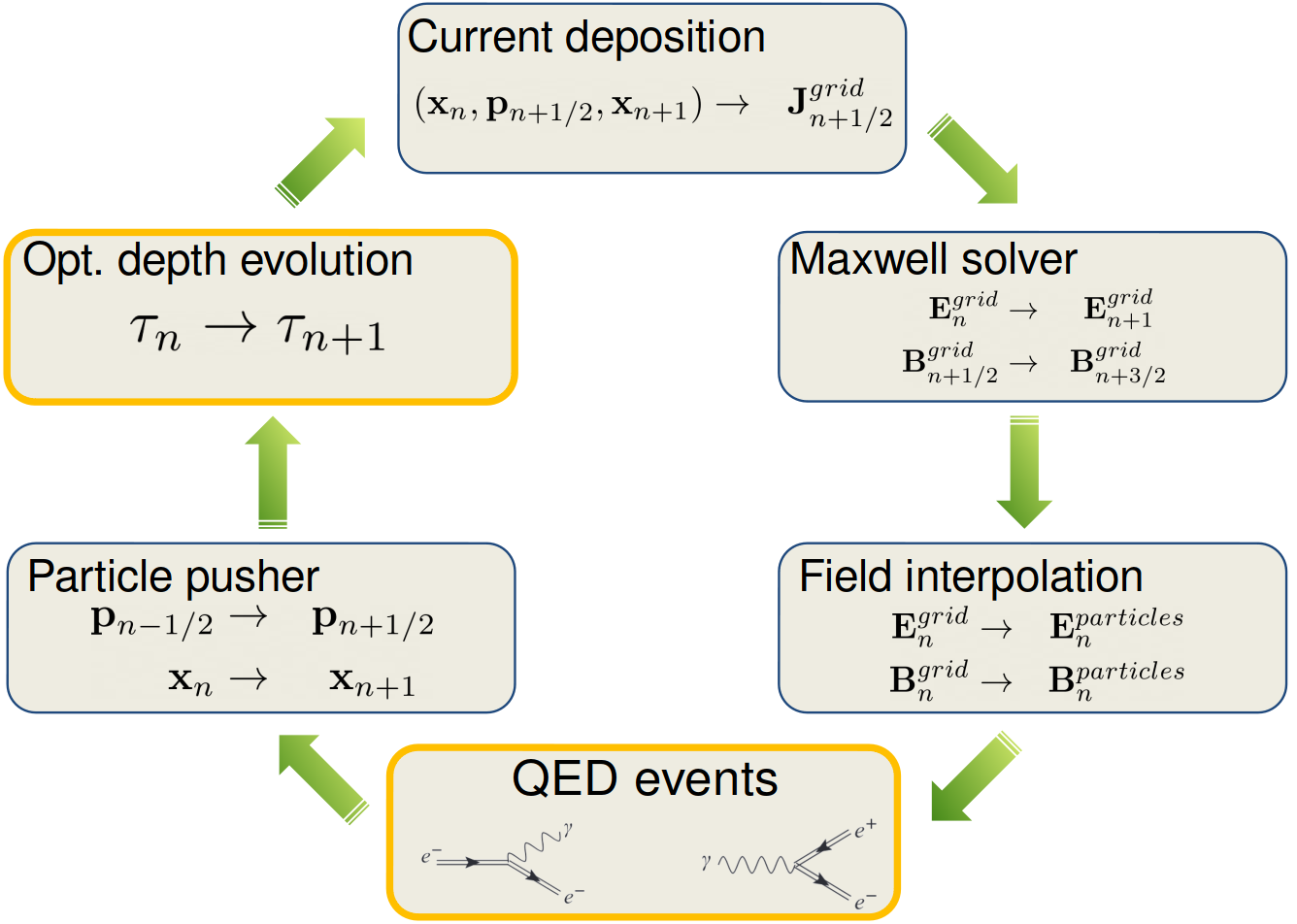}
    \caption{The figure shows the essential sub-steps of a timestep of a PIC code. A gold border marks the sub-steps related to QED effects. In the core PIC algorithm there are two main \emph{actors}: 1) the plasma, which is simulated with a collection of macro-particles (each representing several real particles) and 2) the electromagnetic field, which is simulated on a grid. In each timestep, charged particles move according to the fields interpolated from the grid. The motion of these charged particles generates currents on the grid, which are a source for the evolution of the electromagnetic field. In order to simulate sf-QED two additional sub-steps are added. Particles undergoing QED effects have an additional property, called \emph{optical depth}. This quantity is evolved at each timestep. When the optical depth becomes negative, an event such as Breit-Wheeler pair production or nonlinear synchrotron emission occurs. See \cite{ArberPPCF2015, GonoskovPRE2015} and section \ref{sec:num_meth} for more details.  }
    \label{fig:scheme}
\end{figure}

One of the frontiers of modern physics research deals with physical scenarios where the electromagnetic fields are so strong that Quantum ElectroDynamics (QED) plays a substantial role, specifically in the so-called \emph{strong-field} regime (sf-QED). These scenarios range from the 
interaction region of an ultra-intense laser pulse with a plasma~\cite{MarklundRevModPhys2006, BellPRL2008,RidgersPRL2012,DiPiazzaRevModPhys2012,BulanovPRA2013, PoderPRX2018, ColePRX2018, BucksbaumLOI2020, ZhangPoP2020, GonoskovARXIV2021} to extreme astrophysical
objects, such as pulsar magnetospheres~\cite{BlandfordMNRAS1977, RuffiniPhysRep2010}, black-holes~\cite{CurtisRevModPhys1982, HardingRepProgPhys2006, UzdenskyRepProgPhys2014, PhilippovPRL2020}, or gamma-ray bursts~\cite{MeszarosAJ2001}. Numerical modeling is essential
to gain insights into these scenarios and to assist the experimental investigation of ultra-intense laser-matter interaction. Since a kinetic description of the plasma is usually required, Particle-In-Cell (PIC) codes~\cite{BirdsallBook1985, HockneyBook1988, ArberPPCF2015} are often the numerical tool of choice. Moreover, PIC simulations can include the most relevant sf-QED effects in these regimes~\cite{DuclousPPCF2010, RidgersJCP2014, LobetThesis2015, GonoskovPRE2015, LobetJoP2016}, 
such as the emission of high-energy photons via the \emph{inverse Compton} process~\cite{NikishovJETP1964} (also known as \emph{nonlinear synchrotron emission}) and the decay of a high-energy photon into an electron-positron pair via the \emph{nonlinear Breit-Wheeler pair production} process~\cite{NikishovJETP1964, ErberRevModPhys1966, BaierJETP1968, RitusJSVR1985, KirkPPCF2009} (see Fig. \ref{fig:scheme} for a scheme showing the core algorithms of a PIC code including sf-QED effects).\\
Realistic simulations often demand a substantial amount of computational resources, to the point that the most ambitious numerical campaigns can only be performed on the most powerful High-Performance Computing (HPC) facilities ~\cite{HeldensACM2020}. Most of those machines offload a conspicuous fraction of their calculations to specialized hardware~\cite{VazhkudaiSC2018} (i.e., Graphics Processing Units, GPUs) or make use of CPUs specifically designed for HPC needs~\cite{MitsuhisaIEEE2020}. Few PIC codes in use in the plasma physics community can efficiently take advantage of those machines. Moreover, only a fraction of these codes is distributed as free and open-source software. Requiring the capability to simulate at least the most relevant QED processes further restricts the choice. \\
OSIRIS~\cite{FonsecaICCS2002}, Picador~\cite{BastrakovJCS2012}, VPIC 2.0~\cite{BirdIEEE2021}, and PIConGPU~\cite{PIConGPU2013} are popular codes able to take advantage of modern, GPU-based supercomputers. However, only PIConGPU and VPIC 2.0 are available as free and open-source software. Moreover, while OSIRIS and Picador have very comprehensive QED modules, PIConGPU doesn't implement nonlinear Breit-Wheeler pair production yet, and VPIC 2.0 doesn't implement QED processes. Smilei~\cite{DerouillatCPC2018}, EPOCH~\cite{ArberPPCF2015}, and Tristan-MP v2~\cite{SpitkovskyAIP2005} are well-known open-source, massively parallel PIC codes with comprehensive QED modules, but they are not currently designed to take advantage of GPU-based supercomputers. Calder~\cite{LifschitzJCompPhys2009, LobetThesis2015} is also well-known for its very comprehensive QED modules~\cite{LobetJoP2016, MartinezPoP2019}, but it is not open-source nor optimized for large-scale GPU-based HPC machines. \\
This paper presents PICSAR-QED~\cite{picsarqed_repo}, a module part of the PICSAR library~\cite{PICSAR}, which has been coupled with the WarpX~\cite
{VayNima2018, MyersParComp2021} PIC code in order to simulate sf-QED processes relevant for extreme plasma physics scenarios. WarpX is an open-source code developed within the framework of the Exascale Computing Project~\cite{MessinaIEEE2017} and designed to provide scalable performance on upcoming exascale supercomputers. PICSAR-QED implements the methods needed to model sf-QED processes in PIC codes, and  WarpX+PICSAR-QED is -- to the best of the authors' knowledge -- the first open-source PIC code able to simulate sf-QED processes on large-scale GPU-based supercomputers. \\
PICSAR-QED implements primitives designed to be portable across different architectures (CPUs and GPUs). It's conceived to be easily included in existing projects, and is released as a standalone open-source project, contributing a carefully validated module to the open-source PIC ecosystem, of which other PIC codes not currently implementing sf-QED could take advantage. \\
In this work, we first review the physical models implemented in PICSAR-QED (section~\ref{sec:phys_proc}). Then, we review the numerical methods to model sf-QED processes in PIC codes, we discuss the specific implementation choices made in PICSAR-QED, and we present a detailed validation of the methods provided by the library (section~\ref{sec:num_meth}). In section~\ref{sec:perftest} we show performance benchmarks on different architectures. Finally, we describe in section~\ref{sec:wrpx} the integration of PICSAR-QED in WarpX and present benchmarks with existing results from the literature.

\section{Physical processes implemented in PICSAR-QED}
\label{sec:phys_proc}
The extreme plasma physics scenarios mentioned in the introduction are characterized by electromagnetic fields so strong that relevant QED processes pertain to the strong-field regime of QED. Strong-field here refers to the electromagnetic fields being comparable to the QED critical field $E_s$, also known as \emph{Schwinger field} ~\cite{Sauter1931, Heisenberg1936, SchwingerPR1951}:
\begin{equation}
E_s = \dfrac{m_e^2  c^3}{q_e \hbar} \approx 1.32 \cdot 10^{18}~\textrm{V/m}
\end{equation}
where $m_e$ is the electron mass, $c$ is the speed of light, $q_e$ is the elementary charge and $\hbar$ is the reduced Plank constant.
Such a tremendous field is far beyond current experimental capabilities, being roughly three orders of magnitude higher than the strongest electric fields available on Earth~\cite{YoonOptica2021}. However, it can be approached in the reference frame of ultra-relativistic particles. Indeed, in this case, the actual parameter of interest for sf-QED is the parameter $\chi$, defined as the ratio between the electric field in the reference frame of the particle and $E_s$.  $\chi$ is Lorentz-invariant and is called \emph{quantum parameter}.
For an electron or a positron with 4-momentum $p^\mu$ propagating in a region where the electromagnetic field tensor is $F^{\mu\nu}$, $\chi$ is defined as follows~\cite{RitusJSVR1985} :
\begin{equation}
\chi = \frac{\vert F_{\mu \nu} p^{\nu} \vert}{E_s m_e c} = \frac{\gamma}{E_s} \sqrt{\left({\mathbf{E} +\mathbf{v} \times \mathbf{B}}\right)^2 - \left({\frac{\mathbf{v} \cdot\mathbf{E}}{c}}\right)^2}
\end{equation}
where $\mathbf{E}$ is the electric field, $\mathbf{B}$ is the magnetic field, $\mathbf{v}$ is the velocity of the particle, and $\gamma$ is its Lorentz factor.  
We attain the so-called \emph{full quantum regime} of sf-QED when $\chi > 1$, while strong-field QED effects rapidly vanish for $\chi < 1$.\\
For high-energy photons with 4-momentum $p^\mu$, an analogous Lorentz-invariant $\chi_\gamma$ parameter can be defined:
\begin{equation}
\chi_{\gamma} = \frac{\vert F_{\mu \nu} p^{\nu} \vert}{E_s m_e c} = \frac{\gamma_{\gamma}}{E_s} \sqrt{\left({\mathbf{E} +\mathbf{c} \times \mathbf{B}}\right)^2 - \left({\frac{\mathbf{c} \cdot\mathbf{E}}{c}}\right)^2}
\end{equation}
where $\gamma_\gamma$ is the photon energy normalized to $m_ec^2$ and $\mathbf{c}$ is a velocity vector with a magnitude equal to the speed of light.
The threshold of the full quantum regime of sf-QED is $\chi_\gamma \sim 1$ for photons as well.\\
The physical models used to include the most relevant sf-QED processes in PIC codes -- nonlinear Breit-Wheeler pair production and inverse Compton emission -- are well known from the literature~\cite{NikishovJETP1964, ErberRevModPhys1966, BaierJETP1968, KirkPPCF2009, DuclousPPCF2010, RidgersJCP2014, GonoskovPRE2015, LobetJoP2016, NielPRE2018} and are briefly reviewed below. Schwinger pair production, which is a particularly extreme physical process where the electromagnetic field is strong enough to generate electron-positron pairs from the quantum fluctuations of the vacuum, is also briefly discussed.

\subsection{Nonlinear Breit-Wheeler pair production}
Nonlinear Breit-Wheeler pair production is the decay of a high-energy photon propagating in a strong field into an electron-positron pair. The differential pair production cross section for a photon with quantum parameter $\chi_\gamma$ reads as follows~\cite{ErberRevModPhys1966, LobetThesis2015}:
\begin{equation}
\begin{split}
\dfrac{d^2N}{dt d\chi_-} = &
 \frac{\alpha_{f} m_e c^2}{\pi \sqrt{3} \hbar \gamma_\gamma \chi_\gamma} \int_{X(\chi_\gamma, \chi_-)}^{+\infty} \sqrt{s} K_{1/3}\left({\frac{2}{3} s^{3/2}}\right) ds \\ & - \left[{2 - \chi_\gamma X^{3/2}(\chi_\gamma, \chi_-)}\right] K_{2/3}\left({\frac{2}{3} X^{3/2}(\chi_\gamma, \chi_-)}\right)
 \end{split}
\end{equation}
where $\chi_+$ and $\chi_-$ are respectively the quantum parameter of the emitted positron and of the emitted electron,  $K_{\alpha}$ are the modified Bessel functions of the second kind of order $\alpha$, $\alpha_{f}$ is the fine structure constant, and
\begin{equation}
X(\chi_\gamma, \chi_-) = \left({\dfrac{\chi_\gamma}{\chi_+ \chi_-}}\right)^{2/3}.
\end{equation}
Since particles in very intense electromagnetic fields are usually ultra-relativistic (i.e., $\gamma \gg 1$), the so-called \emph{ultra-relativistic} approximation can be used. In this approximation, the 3-momenta of the product particles are aligned with that of the high-energy photon. Moreover, since $\vert\mathbf{v}\vert \approx c$, the square root terms in the expressions for $\chi$ and $\chi_\gamma$ become almost identical. Finally, since $\gamma \approx \vert \mathbf{p} \vert / {(m_e c)}$, using total momentum conservation we can write:
\begin{equation}
\chi_+ + \chi_- = \chi_\gamma 
\end{equation}
This result is used to replace $\chi_+$ with $\chi_\gamma - \chi_-$, so that 
the differential cross section and the total cross section can be rewritten as follows:
\begin{equation}
\dfrac{d^2N}{dt d\chi_-} = \frac{\alpha_{f} m_e c^2}{\hbar \gamma_\gamma} \chi_\gamma F(\chi_\gamma, \chi_-)
\label{eq:bw_cross_diff}
\end{equation}
\begin{equation}
\dfrac{dN}{dt} = \frac{\alpha_{f} m_e c^2}{\hbar \gamma_\gamma} \chi_\gamma T(\chi_\gamma)
\label{eq:bw_cross_tot}
\end{equation}
where
\begin{multline}
 T(\chi_\gamma) = \int_0^{\chi_\gamma} F(\chi_\gamma, u) du = \\ \dfrac{1}{\pi \sqrt{3}\chi_\gamma^2} \int_0^{\chi_\gamma} \left[{ \int_{X(\chi_\gamma, u)}^{+\infty} \sqrt{s} K_{1/3}\left({\frac{2}{3} s^{3/2}}\right) ds - \left({2 - \chi_\gamma X^{3/2}(\chi_\gamma, u)}\right) K_{2/3}\left({\frac{2}{3} X^{3/2}(\chi_\gamma, u)}\right) }\right] du .
 \label{eq:t_func_0}
\end{multline}
It is noteworthy that very good and simple asymptotic approximations exist for $T(\chi_\gamma)$~\cite{ErberRevModPhys1966}:
\begin{equation}
 T(\chi_\gamma) \quad \sim \quad 0.16 \frac{K^2_{1/3}\left({\dfrac{2}{3 \chi_\gamma}}\right)}{\chi_\gamma} \quad \sim \quad     \begin{cases} 
      \exp\left({-\dfrac{2}{3 \chi_\gamma}}\right) & \chi_\gamma \ll 1 \\
      \chi_\gamma^{-1/3} & \chi_\gamma \gg 1
   \end{cases}
\label{eq:t_func}
\end{equation}
Equation \eqref{eq:bw_cross_tot} allows determining the probability of a photon to decay into an electron-positron pair within a given time interval. Indeed, if the timestep of the simulation is $\Delta t$, this probability is $p_{decay} = 1 - \exp(-\Delta t dN/dt)$. In a Monte Carlo approach, a random number $r_{decay}$ from a uniform probability distribution between zero and one can be drawn and compared with $p_{decay}$: if $r_{decay} < p_{decay}$ Breit-Wheeler pair production occurs for the given photon (or, in a PIC simulation, a macro-photon, that is a numerical particle representing several real photons). In practice, however, a different approach is typically followed~\cite{RidgersJCP2014}, in order to avoid a random number extraction per particle at each iteration. In this approach, each macro-photon has a randomly initialized quantity, called \emph{optical depth}, which is reduced at each time-step according to the total cross section. As soon as this quantity reaches zero, Breit-Wheeler pair production occurs (see section \ref{sec:num_meth} for a more in-depth discussion). \\
The energy of the generated particles can be determined using Eq. \eqref{eq:bw_cross_diff}, by calculating the cumulative probability distribution with respect to $\chi_-$: 
\begin{equation}
P(\chi_\gamma, \chi_-) = \frac{\int_0^{\chi_-} F(\chi_\gamma, u) du }{\int_0^{\chi_\gamma} F(\chi_\gamma, u) du}
\label{eq:bw_prob_dist}
\end{equation}
The quantum parameter of the electron $\chi_-$ can be sampled by solving $P(\chi_\gamma, \chi_-) = r$, where $r$ is a random number drawn from a uniform probability distribution in the range $[0,1]$. The quantum parameter of the positron is then simply $\chi_+ = \chi_\gamma - \chi_-$. In the ultra-relativistic limit, determining the energy and momenta of the generated particles is straightforward if the quantum parameters  $\chi_\pm$ are known. Indeed, their kinetic energy can be calculated as $ K_{\pm} = (\gamma_\gamma - 2) \chi_\pm / \chi_\gamma$.

\subsection{Inverse Compton photon emission}
The inverse Compton photon emission is the emission of a high-energy photon from a charged particle (e.g., an electron or a positron) propagating in a strong electromagnetic field. PICSAR-QED implements the model described in~\cite{RidgersJCP2014}, which is summarized here for completeness (it is worth noting that a slightly different notation is adopted: in~\cite{RidgersJCP2014} $\chi$ is replaced with $\eta$ and $\chi_{\gamma}$ is replaced with $2\chi$).\\
The differential cross-section for the inverse Compton scattering process reads as follows:
\begin{equation}
\dfrac{d^2N}{dt d\chi_\gamma} = \frac{2}{3}\frac{\alpha m_e c^2}{\hbar} \dfrac{1}{\gamma} \frac{  \dfrac{\sqrt{3}}{2 \pi} (\chi_\gamma/\chi) \left[{ \int_{Y(\chi,\chi_\gamma/\chi)}^\infty K_{5/3}( s ) ds  + \frac{(\chi_\gamma/\chi)^2}{1-(\chi_\gamma/\chi)} K_{2/3}(Y(\chi,\chi_\gamma/\chi)) }\right]  }{\chi_\gamma}
\label{eq:qs_diff_cross_exp}
\end{equation}
where 
\begin{equation}
Y(\chi, \chi_\gamma) = \dfrac{2}{3} \dfrac{\chi_\gamma/\chi}{\chi(1-\chi_\gamma/\chi)} .
\end{equation}
Equation \eqref{eq:qs_diff_cross_exp} can be re-written as
\begin{equation}
\dfrac{d^2N}{dt d\chi_\gamma} = \frac{2}{3}\frac{\alpha m_e c^2}{\hbar} \dfrac{1}{\gamma} \dfrac{S(\chi, \xi)}{\xi \chi}
\label{eq:qs_diff_cross}
\end{equation}
where we introduced $\xi = \chi_\gamma/\chi$ . As for Breit-Wheeler pair production, the ultra-relativistic approximation applies. Therefore, since the photon is emitted within a cone of amplitude $\alpha \approx 1/\gamma$, for $\gamma \gg 1$ the photon can be safely considered to be emitted along the direction of the momentum of the emitting particle.
Within the ultra-relativistic approximation, it is also trivial to show that $\xi < 1 $. \\
The total cross section is obtained by integrating Eq. \eqref{eq:qs_diff_cross} over $\chi_\gamma$ from $0$ up to $\chi$:
\begin{equation}
\dfrac{dN}{dt} = \frac{2}{3}\frac{\alpha m_e c^2}{\hbar} \dfrac{1}{\gamma} G(\chi)
\label{eq:qs_total_cross}
\end{equation}
where
\begin{align}
 G(\chi) & = \int_0^{\chi} \dfrac{S(\chi, u/\chi)}{u} du = \\ 
 & \int_0^{1} \dfrac{S(\chi, \xi)}{\xi} d\xi =  \int_0^{1}{ \dfrac{\dfrac{\sqrt{3}}{2 \pi} \xi \left[{ \int_{Y(\chi,\xi)}^\infty K_{5/3}( s ) ds  + \frac{\xi^2}{1-\xi} K_{2/3}(Y(\chi,\xi)) }\right]  }{\xi} d\xi}
 \label{eq:qs_g_func}.
\end{align}
Equation \eqref{eq:qs_g_func} allows determining the probability of an electron or a positron to emit a high-energy photon via inverse Compton emission, with a procedure identical to that described for Breit-Wheeler pair production. 
As for Breit-Wheeler pair production, the quantum parameter $\chi_\gamma$ of the generated photon is determined using the cumulative probability distribution:
\begin{equation}
P(\chi, \xi) = \frac{\int_0^{\xi} S(\chi, u) du }{\int_0^{1} S(\chi, u)}
\label{eq:qs_prob_dist}
\end{equation}
Once $\chi_\gamma$ is known, the energy of the generated photons can be determined trivially using the ultra-relativistic approximation: $\gamma_\gamma = (\gamma-1) \xi$. Finally, the kinetic energy of the emitting particle must be reduced by $\gamma_\gamma m_e c^2$.
\subsection{Schwinger pair production}
Schwinger pair production is the generation of electron-positron pairs from the fluctuations of the quantum vacuum in the presence of a sufficiently strong electromagnetic field.
An expression for the Schwinger pair production rate per unit volume can be found in~\cite{NarozhnyPRA2004}:
\begin{equation} \label{eq:shw}
\dfrac{d^2N}{dt dV} = \dfrac{q_e^2 E_s^2}{4 \pi^2 \hbar^2 c} \epsilon \eta \coth{\left({\dfrac{\pi \eta}{\epsilon}}\right)}\exp{\left({-\dfrac{\pi}{\epsilon}}\right)}
\end{equation}
where $\epsilon = \mathcal{E}/E_s$ and $\eta = \mathcal{H}/E_s$. $\mathcal{E}$ and $\mathcal{H}$ are given by
\begin{equation}
\mathcal{E} = \sqrt{\sqrt{\mathcal{F}^2 + \mathcal{G}^2} + \mathcal{F}} \quad \quad
\mathcal{H} = \sqrt{\sqrt{\mathcal{F}^2 + \mathcal{G}^2} - \mathcal{F}}
\end{equation}
where $\mathcal{F}$ and $\mathcal{G}$ are the invariants of the electromagnetic field and are equal to
\begin{equation}
\mathcal{F} = (\mathbf{E}^2 - c^2 \mathbf{B}^2)/2
\end{equation}
\begin{equation}
\mathcal{G} = c \mathbf{E} \cdot \mathbf{B} 
\end{equation}
Electron-positron pairs generated via the Schwinger pair production process can be initialized at rest. In principle, the electromagnetic field should lose an amount of energy equal to $2 m_e c^2$ when pairs are created via the Schwinger process. However, 
since the field loses significantly more energy while accelerating these particles to relativistic velocities immediately after their creation, the small energy loss due to the rest-mass energy of the pair can be safely disregarded.\\
Schwinger pair production is implemented in PICSAR-QED. However, since the implementation is relatively simple with respect to the other sf-QED process (the pair production rate is not very expensive to compute and product particles are initially at rest), it will not be further discussed.

\section{Numerical implementation}
\label{sec:num_meth}
The total cross-sections for Breit-Wheeler pair production and inverse Compton emission have quite complex expressions, featuring special functions and multiple integrals, which would be too computationally expensive to evaluate for each particle at each time step. Indeed, in the standard PIC algorithm, the number of operations per particle per timestep is relatively small if compared with what would be required to compute QED cross-sections. Therefore, their evaluation at runtime would largely dominate the simulation time, unacceptably slowing down the simulation. For this reason, as documented in the literature~\cite{RidgersJCP2014, LobetThesis2015, LobetJoP2016}, the standard approach is to reformulate the total cross-sections as a product between simple numerical factors and a numerically expensive function, which is pre-computed and stored in a one-dimensional lookup table. For instance, the right hand side of Eq. \eqref{eq:bw_cross_tot} is the product of a constant ($\alpha_f m_e c^2 / \hbar$), the normalized photon energy $\gamma_\gamma$ (which is a simple function of the photon momentum), the quantum parameter $\chi_\gamma$ (which is a simple function of the photon momentum and of the electromagnetic field), and the function $T(\chi_\gamma)$, which contains all the other terms of the total cross section (see Eq. \eqref{eq:t_func_0}). Similarly, for the inverse Compton emission total cross-section, all the numerically expensive terms can be absorbed into the $G(\chi)$ function (see Eq. \eqref{eq:qs_total_cross} and Eq. \eqref{eq:qs_g_func}). The cumulative probability distributions - required to determine the properties of the product particles - are also unpractical to compute at runtime. Therefore, Eq. \eqref{eq:bw_prob_dist} and \eqref{eq:qs_prob_dist} are pre-computed over a finite set of parameters and the result is stored in two-dimensional lookup tables.\\
As mentioned in section \ref{sec:phys_proc}, in principle a random number per particle at each timestep should be drawn in order to determine if a sf-QED process occurs (two underlying assumptions are that the QED cross sections do not vary significantly over one timestep and that the probability of a QED process to occur during a timestep is significantly smaller than one). However, generating pseudo-random numbers can have a significant numerical cost, depending on the algorithm. Therefore, as documented in the literature~\cite{RidgersJCP2014}, the preferred approach is to assign a quantity $\tau$, called \emph{optical depth}, to each particle which may undergo a sf-QED process. $\tau$ is extracted from an exponential probability distribution $p(x) = \exp(-x)$, and at each iteration it is updated as $\tau_{n+1} = \tau_{n} - dN/dt \Delta t$, where $dN/dt$ is the total cross-section of either Breit-Wheeler pair production or inverse Compton photon emission. This second approach is equivalent to the former, but with a reduced computational cost. Moreover, from a numerical point of view, the simpler loop on the particles to update the optical depth offers more opportunities for the compiler to optimize the code (e.g., exploiting \emph{Single Instruction on Multiple Data} parallelization on CPU architectures).\\
This section describes the specific implementation choices made for PICSAR-QED. In particular, we provide details on how lookup tables are calculated and how interpolation within these tables is performed. We also assess how the precision of the lookup tables (number of points and use of single or double precision) affects the accuracy of the results. This is particularly important from the perspective of a user. Indeed, although the general idea of the method to implement sf-QED processes in PIC codes has already been described in the literature, to the best of the authors' knowledge, detailed guidelines on how to choose the parameters of the lookup tables have never been published. At the end of the section, we finally discuss specific choices aimed at achieving portability across multiple architectures.\\
Before delving into the implementation choices made for PICSAR-QED, it is important to clarify that methods to compute the lookup tables are provided for CPU architectures only. This is due to the fact that they require special functions not yet implemented for GPUs (e.g., Bessel functions of fractional order), and rely on the CPU-only library Boost for sophisticated quadrature methods, such as \emph{tanh-sinh}~\cite{MoriQuad1985, BaileyExpMath2005}. Computing the lookup tables typically requires only few tens of seconds on a multi-core CPU, so, in principle, they could be generated at the beginning of each simulation. In practice, it is often more convenient to store them on disk and load them whenever needed (lookup tables typically require only few megabytes of storage). 

\subsection{Nonlinear Breit-Wheeler pair production: implementation choices and benchmarks}
\label{subsec:bw_impl}
\begin{figure}
\includegraphics[width=1.0\columnwidth]{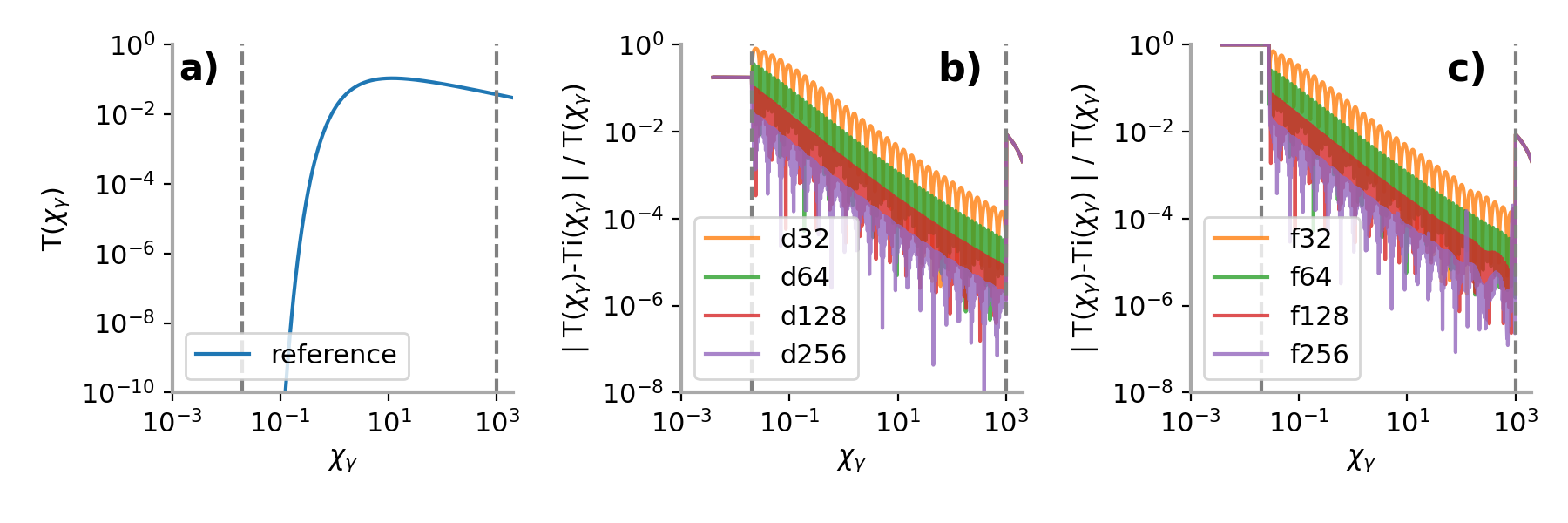}
\caption{\label{fig:bw_dndt_err} a) $T(\chi_\gamma)$ function for a wide range of $\chi_\gamma$. The vertical dashed lines show the limits of the table, i.e. $\chi_{\gamma, min}$ and $\chi_{\gamma, max}$. b) Relative error of $T(\chi_\gamma)$ interpolated from lookup table as a function of the number of table points (the lookup tables were computed in double precision and the interpolation is carried out in double precision as well). c) Same as b), but in single precision.}
\end{figure}

\begin{figure}
\centering
\includegraphics[width=1.0\columnwidth]{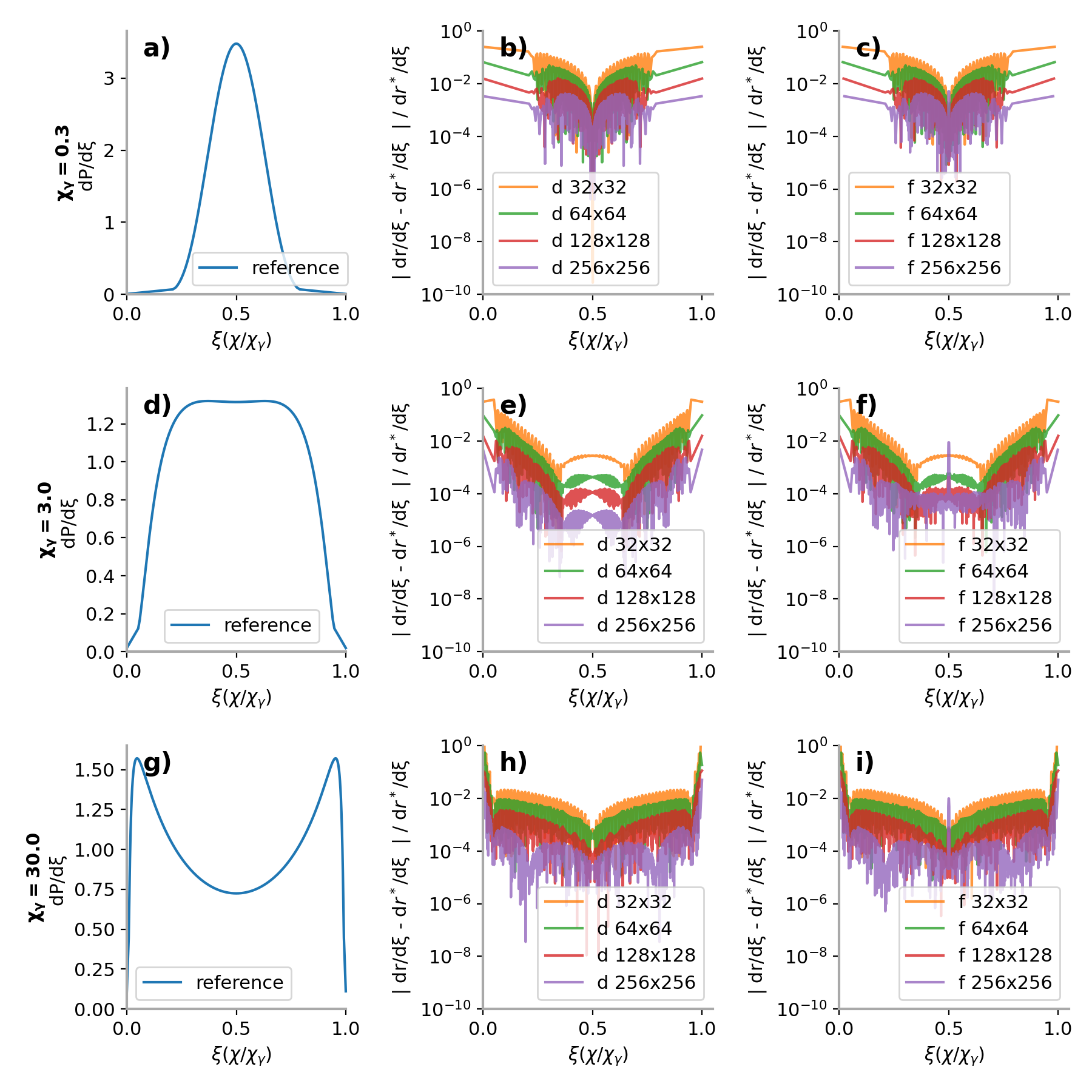}
\caption{\label{fig:bw_pairprod_err}  a,d,g) Probability distribution of the $\xi$ parameter of an electron (or a positron) generated via Breit-Wheeler pair production, for $\chi_\gamma = 0.3, 3.0$ and $30.0 $. b,e,h) Relative error of the 2D lookup table as a function of $N_\gamma \times N_\xi$. c,f,i) Same as b,e,h), but in single precision. The error of the 2D lookup table is calculated as follows: First, using the lookup table $N_\gamma \times N_\xi$, for a given $\chi_\gamma$ we compute $\xi$ as a function of $r$, where $0 \leq r < 1$ ($r$ corresponds to the random number used to extract a $\chi$ parameter from the lookup table). Then, we use Eq. \eqref{eq:qs_prob_dist} to compute $r^* = P(\chi_\gamma, \xi)$ for each $\xi$. We finally define our error as $\dfrac{\vert dr/d\xi - dr^*/d\xi \vert}{dr^*/d\xi}$, where the derivatives are calculated numerically. $dr/d\xi$ is indeed proportional to the number of product particles generated with $\chi = \xi \chi_\gamma$.}
\end{figure}

Two lookup tables are needed for Breit-Wheeler pair production: a one-dimensional table for $T(\chi_\gamma)$ and a two-dimensional table for the cumulative probability distribution $P(\chi_\gamma, \chi)$.
For $T(\chi_\gamma)$, PICSAR-QED library adopts a solution very similar to that described for the Smilei PIC code~\cite{DerouillatCPC2018, LobetThesis2015}. The $T(\chi_\gamma)$ table is generated between a minimum value $\chi_{\gamma, min}$ and a maximum value $\chi_{\gamma, max}$, with $N_{\chi_\gamma}$ points logarithmically distributed between the extrema (actually, $\ln{T}$ is stored in the table). The choice of a logarithmic scale for $\chi_\gamma$ allows spanning several orders of magnitude with a limited number of points, following the strategy proposed in~\cite{RidgersJCP2014}. 
Outside the extrema of the table, we use the approximations in Eq.~\eqref{eq:t_func}. In practice, this is not a significant issue, provided that $\chi_{min} \lessapprox 0.1
$ and $\chi_{max} \gtrapprox 1000$, since the asymptotic limit in Eq.~\eqref{eq:t_func} is a very good approximation (besides, at $\chi_\gamma \sim 0.1$ Breit-Wheeler pair production rapidly becomes negligible). On the other hand, in the range $\chi_{min} < \chi_\gamma < \chi_{max}$, we perform an interpolation. Specifically, in order to calculate $T(\chi^*_\gamma)$, we first individuate two contiguous tabulated values $\chi_{\gamma , n}$ and $\chi_{\gamma, n + 1}$ such that $\chi_{\gamma, n} \leq \chi^*_\gamma < \chi_{\gamma, n + 1} $. Then, we compute $T(\chi^*_\gamma)$ as :
\begin{equation}
    T(\chi^*_\gamma) = \exp\left({ \ln{T_{\gamma, n}}  + \left( \ln\chi^*_\gamma - \ln{\chi_{\gamma, n}} \right) \dfrac{\ln{T_{\gamma, n+1}} - \ln{T_{\gamma, n}}}{\ln{\chi_{\gamma, n+1}} - \ln{\chi_{\gamma, n}}} }\right)
\end{equation}
where $T_{\gamma, n}$ and $T_{\gamma, n + 1}$ are the tabulated values corresponding to $\chi_{\gamma, n}$ and $\chi_{\gamma, n + 1}$. Figure~\ref{fig:bw_dndt_err} provides detailed results on how different choices of $N_{\chi_\gamma}$ and performing all the calculations in single or double precision affects the accuracy of the table. 
As expected, we find that increasing the number of table points reduces the error. In order to achieve an error below few percents for $\chi_\gamma > 0.1$, tables must be calculated with at least 128 points.  Calculating the tables and performing the interpolation in single or in double precision does not seem to affect the final error significantly. \\
For $P(\chi_\gamma, \chi)$, PICSAR-QED adopts a significantly more complex strategy, which partially differs with respect to implementations described elsewhere. First of all, we consider $P(\chi_\gamma, \xi)$,
where $\xi = \chi/\chi_\gamma$, so that $0 \leq \xi \leq 1$. Moreover, we can exploit the symmetry $P(\chi_\gamma, \xi)  = 1 - P(\chi_\gamma, 1 - \xi)$ to store the table only in the range $0 \leq \xi \leq 0.5$. $P(\chi_\gamma, \xi)$ is then generated in the range $\chi_{\gamma, min} \leq \chi_\gamma \leq \chi_{\gamma, max}$, with $N_{\chi_\gamma}$ points logarithmically distributed between the extrema, and in the range $0 \leq \xi \leq 0.5$, with $N_{\xi}$ linearly spaced points. If $\chi_\gamma < \chi_{\gamma, min}$ or $\chi_\gamma > \chi_{\gamma, max}$, we replace $\chi_\gamma$ with either $\chi_{\gamma, min}$ or $\chi_{\gamma, max}$. This means that a user must choose those extrema in such a way that the whole $\chi_\gamma$ range relevant for a given application is included in the table. \\
When a photon with $\chi_\gamma$ decays into an electron-positron pair via Nonlinear Breit-Wheeler pair production, a random number $r$ is extracted from a uniform distribution between 0 and 1, which is used to calculate the quantum parameters of the generated particles. If $r \leq 0.5$, $\chi$ is the quantum parameter of the generated electron. Otherwise $\chi$ is the quantum parameter of the generated positron. In this second case, we replace $r \rightarrow (1-r)$, so as to enforce $0 \leq r \leq 0.5$. At this point, as in the previous case, we individuate the two contiguous tabulated values $\chi_{\gamma,n}$ and $\chi_{\gamma,n+1}$ such that $\chi_{\gamma,n} \leq \chi^*_\gamma < \chi_{\gamma,n+1} $. We can now define $N_{\xi}$ new values $\mathcal{P}_m$:
\begin{equation}
\mathcal{P}_m = P_{n,m} + \left( \ln\chi_{\gamma} - \ln\chi_{\gamma,n} \right) \frac{P_{n+1,m} - P_{n,m}}{\ln\chi_{\gamma,n+1} - \ln\chi_{\gamma,n}}
\end{equation}
where $P_{n,m}$ is the table value corresponding to $\chi_{\gamma, n}$ and $\xi_m$. By performing a binary search, we can find $m^*$ such that:
\begin{equation}
\mathcal{P}_{m^*} \leq r < \mathcal{P}_{m^* + 1} 
\end{equation}
We can finally calculate $\chi$ with a second linear interpolation: 
\begin{equation}
\chi/\chi_\gamma =  \left({ \xi_{m^*} + \left( r - \mathcal{P}_{m^*} \right) \frac{\xi_{m^*+1} - \xi_{m^*}}{\mathcal{P}_{m^* + 1} - \mathcal{P}_{m^*} } }\right)
\end{equation}
Figure \ref{fig:bw_pairprod_err} provides detailed results on how different choices of $N_{\chi_{\gamma}} \times N_{\xi}$ and performing all the calculations in single or double precision affect the accuracy of the table. Also in this case, increasing table resolution results in a better precision of the table and a minimum resolution of 64-128 points in each dimension is required to keep the relative error below few percents. Again, performing the calculations in single precision does not affect these conclusions significantly.

\subsection{Inverse Compton photon emission: implementation choices and benchmarks}
\begin{figure}
\includegraphics[width=1.0\columnwidth]{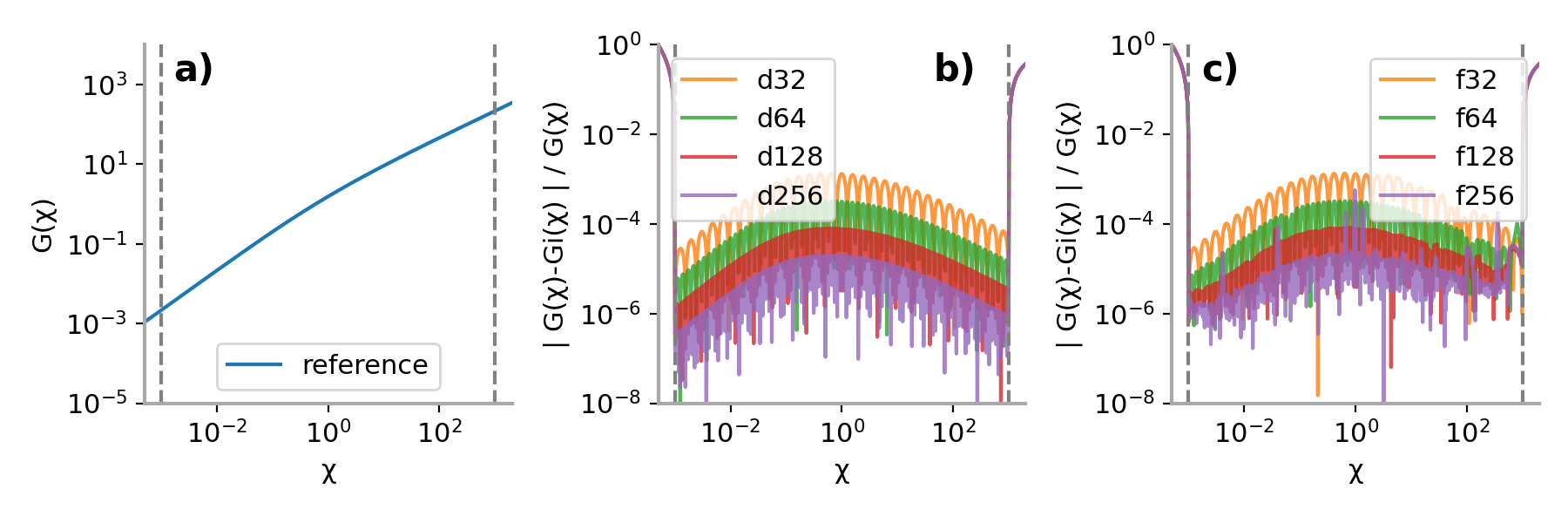}
\caption{\label{fig:qs_dndt_err} a) $G(\chi)$ function for a wide range of $\chi$. The vertical dashed lines show the limits of the table, i.e. $\chi_{min}$ and $\chi_{max}$. b) Relative error of $G(\chi)$ interpolated from the lookup table as a function of the number of table points (double precision). c) Same as b), but in single precision.}
\end{figure}

\begin{figure}[hb!]
\centering
\includegraphics[width=1.0\columnwidth]{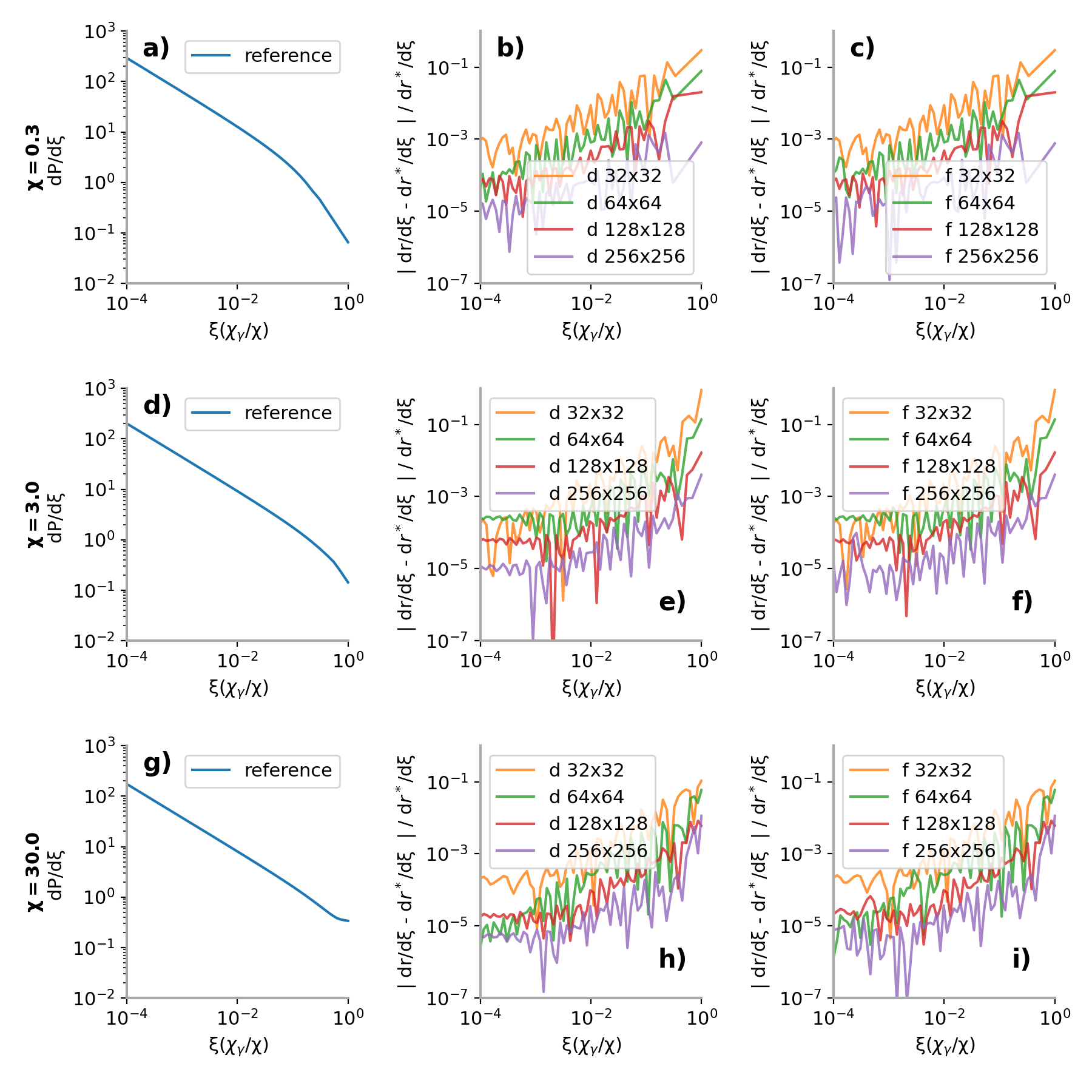}
\caption{
\label{fig:qs_photem_err}  a,d,g) Probability distribution of the $\xi$ parameter of a photon generated via inverse Compton photon emission, for $\chi = 0.3, 3.0$ and $30.0 $. b,e,h) Relative error of the 2D lookup table as function of $N_\gamma \times N_\xi$. c,f,i) Same as b,e,h) but in single precision. }
\end{figure}

As for nonlinear Breit-Wheeler, two lookup tables are needed for inverse Compton photon emission: a one-dimensional table for $G(\chi)$ and a two-dimensional table for the cumulative probability distribution $P(\chi, \chi_\gamma)$.
For $G(\chi)$, the PICSAR-QED library adopts a solution very similar to that described for the Smilei PIC code~\cite{DerouillatCPC2018, LobetThesis2015}. The $G(\chi)$ table is generated between a minimum value $\chi_{min}$ and a maximum value $\chi_{max}$, with $N_{\chi}$ points logarithmically distributed between the extrema (actually, $\ln{G}$ is stored in the table). The choice of a logarithmic scale for $\chi$ allows spanning several orders of magnitude with a limited number of points, following the strategy proposed in \cite{RidgersJCP2014}. Outside the extrema of the table we use either the first or the last value stored in the table, while within this range we perform an interpolation (which means that a user must select those extrema in order to cover all the $\chi$ range of interest). Specifically, in order to calculate $G(\chi^*)$, we first individuate the two contiguous tabulated values $\chi_n$ and $\chi_{n+1}$ such that $\chi_{n} \leq \chi^* < \chi_{n+1} $. Then, we compute $G(\chi^*)$ as :
\begin{equation}
    G(\chi^*) = \exp\left({ \ln {G_{n}}  + \left({ \ln\chi^* - \ln\chi_{n} }\right) \dfrac{\ln {G_{n+1}} - \ln {G_{n}}}{\ln\chi_{n+1} - \ln\chi_{n}} }\right)
\end{equation}
where $G_{n}$ and $G_{n+1}$ are the tabulated values corresponding to $\chi_{n}$ and $\chi_{n+1}$. Figure~\ref{fig:qs_dndt_err} provides detailed results on how different choices of $N_{\chi}$ and performing all the calculations in single or double precision affects the accuracy of the table. As for Breit-Wheeler pair production, increasing the number of table points reduces the error and calculating the tables and performing the interpolation in single or in double precision doesn't seem to affect the final error significantly. However, in this case, a resolution as low as 32 points is already enough to reduce the error below the percent level across the whole $\chi$ range considered here.
As far as $P(\chi, \chi_\gamma)$ is of concern, as for Breit-Wheeler pair production,  PICSAR-QED adopts a significantly more complex strategy, which partially differs with respect to implementations described elsewhere. First of all, we consider $P(\chi, \xi)$,
where $\xi = \chi_\gamma/\chi$, so that $0 \leq \xi \leq 1$. $P(\chi, \xi)$ is then generated in the range $\chi_{min} \leq \chi \leq \chi_{max}$, with $N_{\chi}$ points logarithmically distributed between the extrema, and in the range $\xi_{min} \leq \xi \leq 1$, with $N_{\xi}$ logarithmically distributed points. $\xi_{min}$ must be low enough that photons below the threshold contribute negligibly to the total energy loss via inverse Compton. If $\chi < \chi_{min}$ or $\chi > \chi_{max}$, we replace $\chi$ with either $\chi_{min}$ or $\chi_{max}$, so that the extrema must be selected in order to include all the $\chi$ range of interest. In the table we actually store $\ln P$ instead of $P$.\\
When an electron or a positron with quantum parameter $\chi$ emits a high-energy photon via inverse Compton process, a random number $r$ is extracted from a uniform distribution between 0 and 1. At this point, as in the previous case, we individuate the two contiguous tabulated values $\chi_{n}$ and $\chi_{n+1}$ such that $\chi_{n} \leq \chi^* < \chi_{n+1} $. We can now define:
\begin{equation}
\ln{ \mathcal{P}_m } = \ln { P_{n,m} } + \left( \ln\chi_{\gamma} - \ln\chi_{\gamma,n} \right) \frac{\ln { P_{n+1,m} }- \ln { P_{n,m} } }{\ln\chi_{\gamma,n+1} - \ln\chi_{\gamma,n}}
\end{equation}
where $P_{n,m}$ is the table value corresponding to $\chi_{n}$ and $\xi_m$. By performing a binary search, we can find $m^*$ such that:
\begin{equation}
\ln{\mathcal{P}_{m^*}} \leq \ln { r} < \ln{\mathcal{P}_{m^* + 1}} 
\end{equation}
We can finally calculate $\chi_\gamma$ with a second linear interpolation: 
\begin{equation}
\chi_\gamma = \chi \exp { \left( { \ln \xi_{m^*} + \left( \ln r - \ln\mathcal{P}_{m^*} \right) \frac{\ln \xi_{m^*+1} - \ln \xi_{m^*}}{\ln \mathcal{P}_{m^* + 1} - \ln\mathcal{P}_{m^*} }  } \right) }
\end{equation}
Figure \ref{fig:qs_photem_err} provides detailed results on how different choices of $N_{\chi} \times N_{\xi}$ and performing all the calculations in single or double precision affect the accuracy of the table. As for Breit-Wheeler pair production, increasing the number of table points reduces the error and calculating the tables and performing the interpolation in single or in double precision doesn't seem to affect the final error significantly. A resolution of 128 points in each dimension is required to reduce the error to the few percent level across the whole $\chi$ range considered here.

\subsection{Portability across different architectures}
From a technical point of view, PICSAR-QED is a \verb|C++14|, \emph{header-only} library, designed to integrate easily into other projects, and to provide methods able to run efficiently on different computing architectures. This section describes how these goals are achieved.
\subsubsection*{Support of different unit systems - }
Internal calculations in PICSAR-QED are performed adopting \emph{Heaviside-Lorentz units} with 1 MeV chosen as the reference energy. However, the interface of PICSAR-QED also supports \emph{SI} units and normalized units where the speed of light, the elementary charge, and the electron mass are equal to one and either a reference length or a reference frequency is used (in these cases the value of the reference quantity must be provided). In practice, since the library is written in \verb|C++|, this is achieved via templates, in order to avoid code duplication. The choice of the units is performed at compilation time, in order to avoid overheads at runtime. Since lookup tables are adimensional, once generated they can be used with any choice of unit system.
\subsubsection*{Support for single and double precision - } We provide simple and double precision versions of each method (in practice, since the library is written in \verb|C++| this is also achieved via templates). This flexibility is crucial, especially for methods that should be used at runtime, since on several architectures running in single precision leads to large speedups. Moreover, it allows integrating PICSAR-QED with codes using single precision, avoiding expensive runtime floating point casts. Lookup tables can be computed either in single or double precision, with the former case being significantly faster due to relaxed tolerance required for numerical integration. PICSAR-QED also provides an option to compute the lookup tables in double precision and store them in single precision.
\subsubsection*{Avoid dependencies on specific pseudo-random number generators - }
PICSAR-QED is a Monte Carlo module. Therefore, pseudo-random numbers are needed at runtime. Since pseudo-random number generators have different interfaces in different libraries and performance portability frameworks, we decided not to force the use of a specific pseudo-random number generator, nor to include a pseudo-random number generator in PICSAR-QED. Our design specifications requires that a random number (uniformly distributed between zero and one) is passed to each runtime function requiring randomness. This gives complete freedom to the users on how to generate such random numbers.
\subsubsection*{Compatibility with different architectures - }
PICSAR-QED provides a collection of methods that can be divided into two categories: \emph{runtime} methods (which are actually needed during a Particle-In-Cell simulation) and \emph{lookup table generation methods}, which are needed only to generate lookup tables for later use. Only \emph{runtime} functions need to be portable on different architectures, while methods for lookup-table generation need only to run on CPUs (moreover, their compilation for some architectures, namely GPUs, is not currently possible).
In order to achieve portability across different architectures, all the \emph{runtime} methods are pre-pended with some macros, whose values must be set appropriately to compile the code for CPUs or GPUs, or to use performance portability frameworks like Kokkos~\cite{EdwardsKokkos2014} and AMReX~\cite{ZhangOJ2019}, as explained in detail in appendix A. While Kokkos is primarily designed as a performance portability framework, AMReX is actually a library designed to support massively parallel block-structured adaptive mesh refinement (AMR) applications, but it also offers features enabling performance portability of the applications built on top of it, like the PIC code WarpX.   \\
Another key concept enabling portability concerns the data structures. In this regard, PICSAR-QED provides containers, such as those used internally for the lookup-tables, which must be initialized with methods running on CPUs, while at the same time being available in GPU kernels, if the library is compiled for those architectures. Achieving this may require some effort from the user, but the amount of effort is minimal for Kokkos and AMReX, as well as for for programming models like CUDA~\cite{NickollsACM2008}, as shown in appendix B.

\section{Performance Benchmarks on different architectures}
\label{sec:perftest}
Since PICSAR-QED can be compiled for different architectures and integrated with different performance portability frameworks, we have carried out extensive performance benchmarks of the four most important kernels of the
library: 
\begin{itemize}
\item Breit-Wheeler optical depth evolution
\item Breit-Wheeler pair production
\item Inverse Compton optical depth evolution
\item Inverse Compton photon emission
\end{itemize}
Those benchmarks were carried out with a test program using CUDA on an NVIDIA Quadro GV100 GPU, with a test program using OpenMP on a dual-socket machine with Intel Xeon Gold 6152 CPUs and on an AMD EPYC 7302 CPU, and with a test program using the Kokkos library on all the aforementioned architectures. In some selected cases, the effect of changing the number of threads and enabling non IEEE-compliant aggressive floating-point optimizations (``fast math'') was tested as well. We also performed some initial benchmarks on the Fujitsu A64FX CPU, which demonstrate that PICSAR-QED can be used on this architecture, but the results are too preliminary to be included in a fair benchmark. In all cases the benchmark was carried out with $10^8$ particles, each one having ten real components: the three components of the momentum, the six components of the electromagnetic field, and the optical depth. Those quantities are initialized randomly, drawing each component of $\mathbf{E}$ in $[-E_s/100, E_s/100]$, each component of $\mathbf{B}$ between $[-E_s/(100 c), E_s/(100 c)]$, each component of the momentum in $[-1~GeV/c, 1~GeV/c]$ and the optical depth from an exponential distribution. 1D lookup tables were generated with 256 points, while 2D lookup tables were generated with $256 \times 256$ points. $\chi_{min}$ and $\chi_{\gamma, min}$ were chosen to be, respectively, 0.001 and 0.02. $\chi_{max}$ and $\chi_{\gamma, max}$ were both $10^3$, while $\xi_{min}$ was $10^{-12}$. Each kernel was tested in double precision and in single precision for all the $10^{8}$ particles. Appendix C provides details on how the code was compiled in each case. Table \ref{tab:comp} reports the results of these benchmarks.
\begin{table}
\begin{tabular}{|l|l|c|c|c|c|}
\hline 
{\bf Hardware} & {\bf Test case} & \multicolumn{4}{c}{{\bf Kernel timing} (ms)} $\left({\thead{ \textrm{double} \\  \textrm{float}}}\right)$  \vline \\
 & & BW evol & BW prod & QS evol & QS em \\ 
\hline 
NVIDIA Quadro GV100 & CUDA & \thead{14.52 \\  7.43} & \thead{29.94\\  15.70}  & \thead{14.62 \\  7.39} & \thead{26.88 \\  14.15}  \\ 
\hline
NVIDIA Quadro GV100 & Kokkos+CUDA  & \thead{15.11 \\  7.91} & \thead{109.87 \\  81.44} & \thead{15.11 \\  8.96} & \thead{115.22 \\ 79.87}  \\ 
\hline
2x Intel Xeon Gold 6152 (11) & OpenMP (+precise) &  \thead{305.70 \\  122.51 } & \thead{ 1354.27 \\  948.82} &\thead{323.09 \\  135.89} &\thead{1454.12 \\  973.01} \\ 
\hline
2x Intel Xeon Gold 6152 (11) & Kokkos+OpenMP &  \thead{1027.38 \\  1176.31 } & \thead{ 6622.17 \\  6609.26} &\thead{1194.59 \\  826.53} &\thead{7078.74 \\ 6925.94 } \\  
\hline
2x Intel Xeon Gold 6152 (22) & OpenMP (+precise) &  \thead{144.28 \\  69.70 } & \thead{ 738.31 \\  533.57} &\thead{173.04 \\  68.94} &\thead{789.34 \\  544.17} \\ 
\hline
2x Intel Xeon Gold 6152 (22) & Kokkos+OpenMP &  \thead{531.24 \\  622.76 } & \thead{ 4349.26 \\  4389.06} &\thead{612.26 \\  436.75} &\thead{4582.44 \\  4492.05} \\  
\hline
2x Intel Xeon Gold 6152 (44) & OpenMP (+precise) &  \thead{125.17 \\  57.82 } & \thead{ 452.58 \\  328.04} &\thead{130.38 \\  66.69} &\thead{478.98 \\  342.09} \\ 
\hline
2x Intel Xeon Gold 6152 (44) & Kokkos+OpenMP &  \thead{324.07 \\  369.82 } & \thead{ 3500.50\\  3543.78} &\thead{375.32 \\  263.37} &\thead{3655.18 \\  3606.69} \\ 
\hline
2x Intel Xeon Gold 6152 (88) & OpenMP (+precise) &  \thead{157.50 \\  62.36 } & \thead{ 326.59 \\  243.98} &\thead{135.68 \\  70.14} &\thead{343.62 \\  254.66} \\ 
\hline
2x Intel Xeon Gold 6152 (88) & Kokkos+OpenMP &  \thead{205.15 \\  215.60 } & \thead{ 1135.29 \\  1124.45} &\thead{225.32 \\  166.59} &\thead{1201.02 \\  1093.64} \\ 
\hline
AMD EPYC 7302 (32) & OpenMP &  \thead{338.77 \\ 189.56 } & \thead{ 945.99 \\  926.39} &\thead{401.09\\ 331.63} &\thead{931.46 \\  924.30} \\ 
\hline
AMD EPYC 7302 (32) & OpenMP (+fast math) & \thead{349.69 \\  169.36 } & \thead{ 663.94 \\  558.94} &\thead{357.94 \\  253.24} &\thead{634.82 \\  504.24} \\ 
\hline
AMD EPYC 7302 (32) & Kokkos+OpenMP &  \thead{ 349.17 \\  330.48 } & \thead{ 5242.70 \\  4699.89} &\thead{392.12 \\  248.85 } &\thead{ 4967.67\\  4632.18} \\
\hline
\end{tabular}
\caption{
Performance benchmarks of the main functions provided by PICSAR-QED on different architectures with different paradigms. Each cell reports two numbers, which represent the time required to complete the test in milliseconds (each number is actually the average of three runs of the test). The top number refers to the case in double precision, while the bottom number refers to the case in single precision. For the Intel Xeon Gold 6152 and AMD EPYC 7302 cases, the number in parenthesis is the number of OpenMP threads used for the test. \emph{(+fast math)} and \emph{(+precise)}, as documented in Appendix C, refer to the use of floating point models not strictly IEEE-compliant in the compilation of the code, which normally allows for additional optimizations. 
}\label{tab:comp}
\end{table}
The results show remarkable performances with the NVIDIA GV100 GPU, and a near perfect halving of the time spent in each kernel passing from double precision to single precision (as for many architectures, with NVIDIA GV100 the theoretical maximum FLOPS in single precision is twice the maximum FLOPS attainable in double precision). On the other hand, we rarely observe such halving in the case of CPUs. However, this can be explained as follows. With CPUs, the performance doubling passing from double precision to single precision occurs normally only if a kernel can take full advantage of SIMD vectorization. For SIMD vectorization to occur, a kernel must respect relatively stringent requirements. Pair production and photon emission kernels are too complex to be fully vectorized, especially due to the branching conditions in the binary search step included in those kernels. Optical depth evolution kernels are significantly easier to vectorize, although, as shown in the table, this requires to relax floating point arithmetic (relaxing strict IEEE compliance for floating point operations allows most compilers to perform more extensive optimizations, such as reordering of commutative math operations and using approximate versions of some functions). We also observed that using Kokkos introduces an overhead, which is small (even negligible in some cases) for the optical depth evolution kernels. However, we observed a significant overhead in the case of the pair production kernel and of the photon emission kernel. Investigation of this issue is currently in progress. In any case, although the overhead for some kernels is substantial, in a complete PIC simulation this is not likely to be a significant issue. Indeed, while the optical evolution kernels are executed at each timestep for each particle, pair production and photon emission kernels are relevant only for the few of them undergoing a QED process in a given timestep (for the simulation to be reliable, the timestep must be chosen small enough for pair production and photon emission to occur significantly less than every timestep for each particle~\cite{RidgersJCP2014}). \\
A fairer comparison between different computing architectures requires to take into account their different power consumption (i.e., 250 W for NVIDIA Quadro GV100, 280 W for a pair of Intel Xeon Gold 6152, and 155 W for AMD EPYC 7302). Therefore, table \ref{tab:comp_en} reports, for a selection of cases, the total energy required to apply each kernel to all the particles. We note that, without using Kokkos, performing a test on a NVIDIA Quadro GV100 GPU requires one order of magnitude less energy than performing the same test on CPU. 

\begin{table}
\begin{tabular}{|l|l|c|c|c|c|}
\hline 
{\bf Hardware} & {\bf Test case} & \multicolumn{4}{c}{{\bf Used energy per kernel} (J)} $\left({\thead{ \textrm{double} \\  \textrm{float}}}\right)$  \vline \\
 & & BW evol & BW prod & QS evol & QS em \\ 
\hline 
NVIDIA Quadro GV100 & CUDA & \thead{3.63 \\  1.86} & \thead{7.49\\  3.93}  & \thead{3.66 \\  1.85} & \thead{6.72 \\  3.54}  \\ 
\hline
NVIDIA Quadro GV100 & Kokkos+CUDA  & \thead{3.78 \\  1.98} & \thead{27.47 \\  20.36} & \thead{3.78 \\  2.24} & \thead{28.81 \\ 19.97}  \\ 
\hline
2x Intel Xeon Gold 6152 (88) & OpenMP (+precise) &  \thead{44.1 \\  17.46 } & \thead{ 91.45 \\  68.31} &\thead{37.99 \\  19.64} &\thead{96.21 \\  71.30} \\ 
\hline
2x Intel Xeon Gold 6152 (88) & Kokkos+OpenMP &  \thead{57.44 \\  60.37 } & \thead{ 317.88 \\  314.85} &\thead{63.09 \\  46.65} &\thead{336.29 \\  306.22} \\ 
\hline
AMD EPYC 7302 (32) & OpenMP (+fast math) & \thead{54.20 \\  26.25 } & \thead{ 102.91 \\  86.64} &\thead{55.48 \\  39.25} &\thead{98.40 \\  78.16} \\ 
\hline
AMD EPYC 7302 (32) & Kokkos+OpenMP &  \thead{ 54.12 \\  51.22 } & \thead{ 812.62 \\  728.48} &\thead{60.78 \\  38.57 } &\thead{ 769.99 \\  717.99} \\
\hline
\end{tabular}
\caption{
Performance benchmarks of the main functions provided by PICSAR-QED on different architectures with different paradigms. The table is based on a selection of the data reported in table \ref{tab:comp}. The time measured for each test case was multiplied by the \emph{Thermal Design Point} power of each processor, giving the energy required to complete each test. The top number refers to the case in double precision, while the bottom number refers to the case in single precision. For the Intel Xeon Gold 6152 and AMD EPYC 7302 cases, the number in parenthesis is the number of OpenMP threads used for the test. \emph{(+fast math)} and \emph{(+precise)}, as documented in Appendix C, refer to the use of floating point models not strictly IEEE-compliant in the compilation of the code, which normally allows for additional optimizations. 
}\label{tab:comp_en}
\end{table}

\section{Integration of PICSAR-QED with WarpX PIC code}
\label{sec:wrpx}
In order to integrate PICSAR-QED with WarpX, we followed standard procedures described in the literature~\cite{GonoskovPRE2015}. We first added photon macro-particles. Then, for particles involved in QED processes, we added a new real component: the \emph{optical depth}. We then modified the particle evolution routines in order to update this component for particles participating in QED processes. Only for particles whose optical depth reaches zero, we call either a routine to generate a pair, if the particle is a photon, or a routine to emit a photon, if the particle is an electron or a positron. In order to avoid the generation of too many low-energy photon particles, WarpX allows setting a threshold for keeping such particles after creation. If the photon energy is below such threshold, the emitting particle is slowed down as if the photon were emitted, but the photon particle is discarded immediately after creation. By default, the threshold is set to $2 m_e c^2$, because particles with lower energy will almost certainly not decay into a pair. Since simulating QED effects makes sense only when QED cross-sections are sufficiently high, WarpX offers the possibility to select a minimum quantum parameter threshold below which Breit-Wheeler pair production is ignored and photon emission is simulated with a classical model~\cite{TamburiniNJP2010, VranicCPC2016}. When WarpX is compiled for GPUs, all the routines dealing with QED processes (optical depth initialization and evolution, pair production and photon emission) are carried out in GPU kernels, with the exception of the initial lookup tables generation. Lookup tables generation typically requires a few minutes, and can benefit from multi-core parallelization (which can bring the time required to compute the tables down to few tens of seconds). The user can either generate tables by setting the parameters in WarpX input file or load a lookup table generated beforehand.\\
\subsection{Benchmarks with existing literature}
\begin{figure}
\includegraphics[width=1.0\columnwidth]{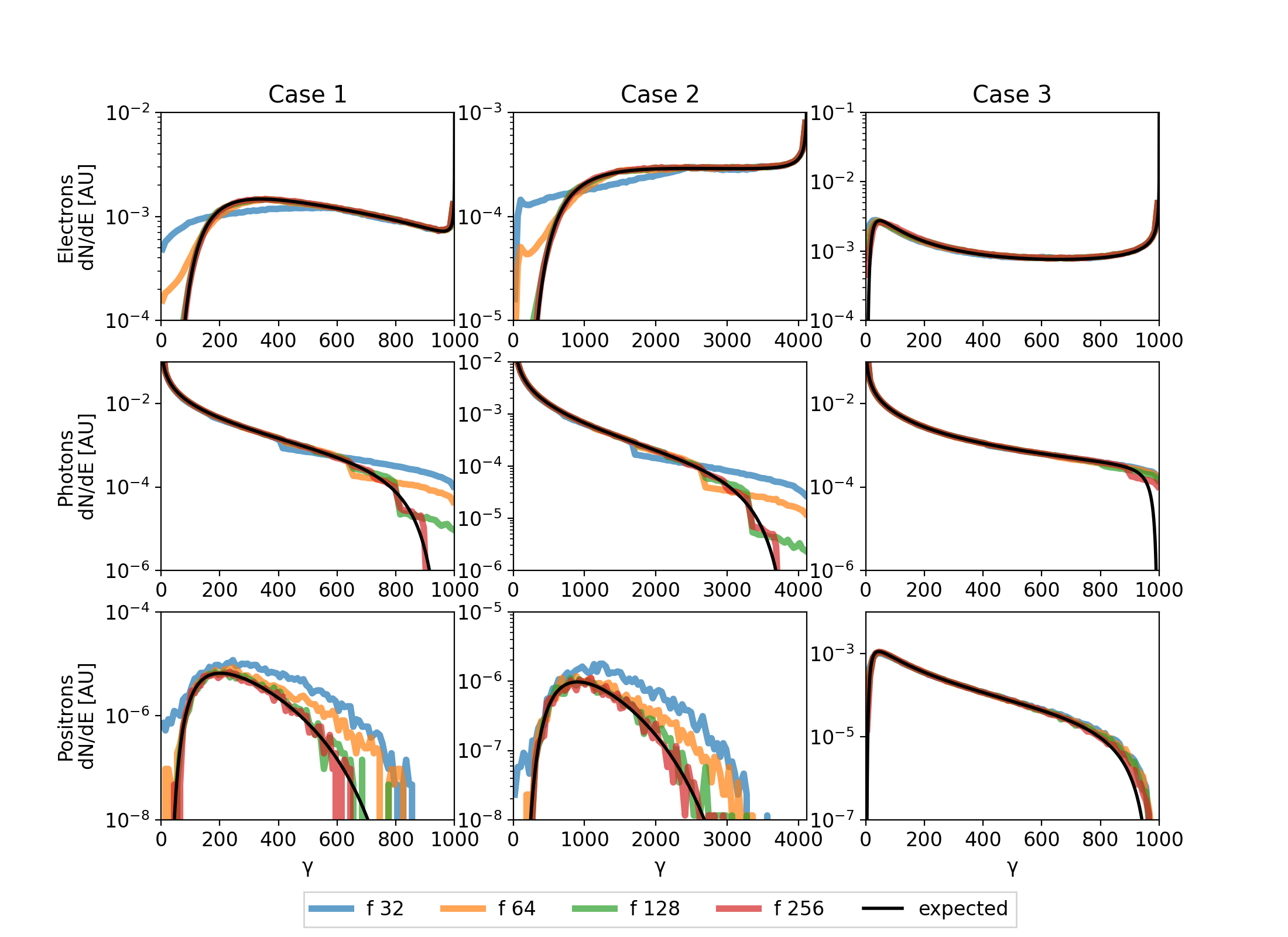}
\caption{\label{fig:phys_bench} The plots show particle energy spectra obtained with WarpX+PICSAR-QED for the 3 simulation cases described in~\cite{RidgersJCP2014}. For each case the simulation was performed multiple times, in single and double precision and using lookup tables of different sizes. Only simulations performed in single precision are shown. In each plot, the black line is the expected result, obtained by directly solving the equations for the evolution of the distribution functions of particles and photons, as described in~\cite{SokolovPRL2010, RidgersJCP2014}.}
\end{figure}

\begin{figure}
\centering
\includegraphics[width=0.6\columnwidth]{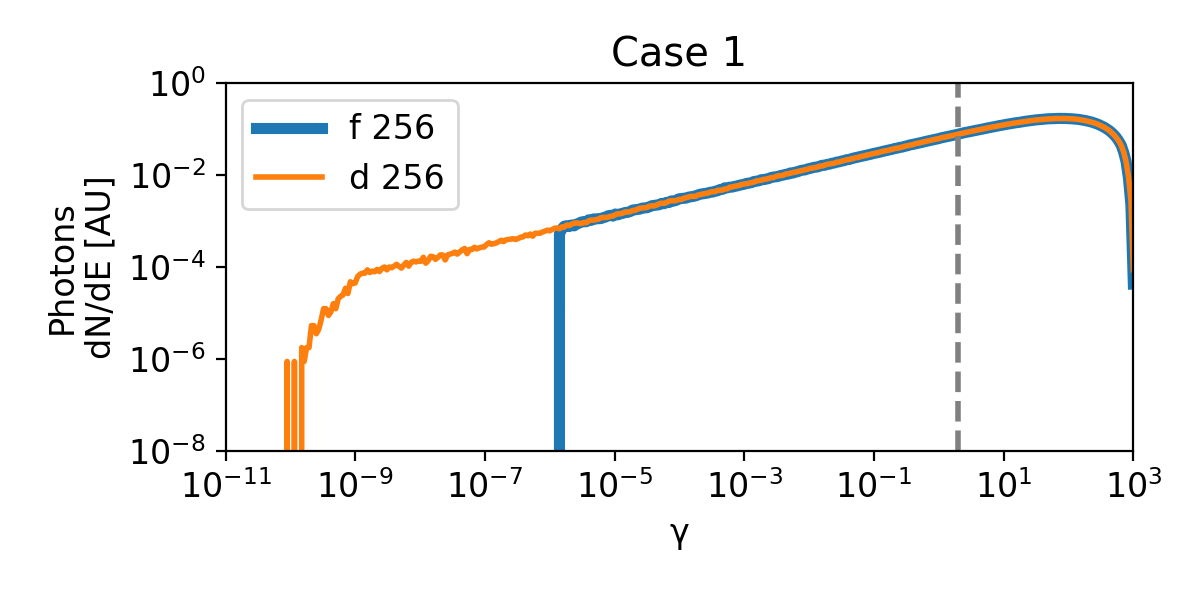}
\caption{\label{fig:phys_bench_2} Photon energy spectra obtained with WarpX+PICSAR-QED for \emph{case 1} described in~\cite{RidgersJCP2014}. The plot shows results obtained in single and double precision, with lookup tables having 256 points in each direction. 
The vertical line marks the $2 m_e c^2$ energy threshold below which photons are deleted by default right after creation.}
\end{figure}

In order to test WarpX+PICSAR-QED, we reproduced the three test cases mentioned in ~\cite{RidgersJCP2014}. In this paper by Ridgers et al., the authors describe their implementation of a QED module to simulate QED processes in laser-plasma interaction. They test this module (which is built-in in the EPOCH PIC code) in the following three physical scenarios:
\begin{enumerate}
\item an electron population with initial Lorentz factor $\gamma_0 = 1000$ propagating in a perpendicular, static, magnetic field $B_0 = 10^{-3} E_s/c$. This case corresponds to an initial quantum parameter $\chi = 1$. The duration of the simulation is 1 fs.
\item an electron population with initial Lorentz factor $\gamma_0 = 4120$ counter-propagating with a circularly polarized plane wave with $E_0 = 1.22 \times 10^{-4} E_s$. This case corresponds to an initial quantum parameter $\chi = 1$. The duration of the simulation is 3 fs.
\item an electron population with initial Lorentz factor $\gamma_0 = 1000$ propagating in a perpendicular, static, magnetic field $B_0 = 9 \times 10^{-3} E_s/c$. This case corresponds to an initial quantum parameter $\chi = 9$. The duration of the simulation is 0.1 fs.
\end{enumerate}
We simulated these three cases with WarpX+PICSAR-QED, using the code in single precision and with lookup tables resolutions ranging from 32 points to 256 points in each dimension. Fig. \ref{fig:phys_bench} reports the energy spectra of the particles at the end of the simulation, for the three cases and for the 4 lookup table resolutions tested in the benchmark. Our results closely reproduce those published in the paper by Ridgers et al., which confirms the correctness of the QED modules in WarpX.
Concerning the results reported in Fig. \ref{fig:phys_bench}, we repeated the tests also in double precision, observing only minor differences. Running the code in single precision (and therefore using lookup tables calculated in single precision), however, has an effect on the photon energy spectrum at very low energy, as shown in fig. \ref{fig:phys_bench_2}. The spectra are very similar, but in single precision the low energy part of the spectrum is truncated. This is simply due to the smaller range of real numbers that single precision can represent, and is a minor concern, since photons with such a low energy have a negligible impact on the total radiative losses. Moreover, those photons are orders of magnitude below $2 m_e c^2$, which means that they cannot decay into electron-positron pairs via Breit-Wheeler pair production.

\section{Conclusions}
This paper presents PICSAR-QED, a Monte Carlo module to simulate strong-field QED processes in PIC codes designed for next-generation supercomputers. PICSAR-QED is conceived to be easily included in other processes and to be portable to different architectures. The outcome of several benchmarks, which demonstrate that the library produces reliable results and can run efficiently on CPUs and GPUs, is reported. Moreover, a detailed investigation on how the parameters used to generate the lookup tables affect the simulation results is provided. The main outcomes of this investigation are that (i) performing the calculations of the QED routines in single precision does not affect the accuracy of the results, and (ii) lookup tables with 128 or 256 points in each direction are sufficient to ensure negligible interpolation errors. Finally, the integration of PICSAR-QED with a state-of-the-art PIC code, WarpX, and the validation of such integration are discussed. WarpX+PICSAR-QED has been already used for production simulations of laser-plasma interaction at QED relevant intensities~\cite{FedeliPRL2021}.

\section*{Appendix A: portability across different architectures }
In PICSAR-QED, all the runtime methods are pre-pended with two macros, \verb+PXRMP_GPU_QUALIFIER+ and \verb+PXRMP_FORCE_INLINE+, as exemplified here
for the function calculating the Schwinger pair production rate:
\begin{Verbatim}[samepage=true]
    template<
        typename RealType,
        unit_system UnitSystem = unit_system::SI
    >
    PXRMP_GPU_QUALIFIER
    PXRMP_FORCE_INLINE
    RealType pair_production_rate(
        const RealType ex, const RealType ey, const RealType ez,
        const RealType bx, const RealType by, const RealType bz,
        const RealType ref_quantity = math::one<RealType>);
\end{Verbatim}

By setting those two macros according to table \ref{tab:qual}, runtime functions can be either:
\begin{itemize}
    \item compiled directly for CPUs.
    \item compiled directly for NVIDIA GPUs using CUDA.
    \item made compatible with the Kokkos performance portability library (and then compiled for CPUs or GPUs).
    \item made compatible with the AMReX library (and then compiled for CPUs or GPUs).
\end{itemize}

\SaveVerb{vrba}+PXRMP_GPU_QUALIFIER+
\SaveVerb{vrbb}+PXRMP_FORCE_INLINE+
\SaveVerb{vrb}|inline __attribute__((always_inline))|
\begin{table}[h]
\begin{tabular}{|c|c|c|}
\hline 
{\bf Case} & \verb+PXRMP_GPU_QUALIFIER+ & \verb+PXRMP_FORCE_INLINE+ \\ 
\hline 
Serial or openMP (CPU) & - & compiler-dependent$^{*}$  \\ 
\hline 
CUDA &   \verb+__host__ __device__+ & \verb+__forceinline__+ \\ 
\hline 
Kokkos & - & \verb+KOKKOS_FORCEINLINE_FUNCTION+ \\ 
\hline 
AMReX & \verb+AMREX_GPU_HOST_DEVICE+ & \verb+AMREX_FORCE_INLINE+ \\ 
\hline 
\end{tabular}
\caption{
Reference table on how to set \protect{\UseVerb{vrba}} and \protect{\UseVerb{vrbb}} to compile PICSAR-QED methods for CPUs and GPUs and/or to make them compatible with AMReX or Kokkos. \\ $^*$ for most compilers this must be \protect{\UseVerb{vrb}}, which is the default value in PICSAR-QED. \label{tab:qual}}
\end{table}

\section*{Appendix B: details on the data structures used in PICSAR-QED}
This appendix provides some details on the underlying data structures used for the lookup tables by PICSAR-QED. Lookup tables use several 1D vectors to store table data (even 2D lookup tables use 1D vectors as their underlying data structure).
Those vectors must be manipulated on CPU when the lookup table is generated or loaded from disk, but they must be accessible within GPU kernels if PICSAR-QED is compiled for GPU architectures.
Moreover, a machinery to seamlessly pass data between the CPU and the GPU is needed. AMReX already provides such a vector type: \verb+amrex::Gpu::DeviceVector+ (with the additional prescription of calling a GPU synchronization method after table initialization). In pure CUDA, a very thin wrapper around the \verb+device_vector+ provided by CUDA thrust library is enough to fulfill these requirements: 
\begin{verbatim}
template<typename RealType>
class ThrustDeviceWrapper : public thrust::device_vector<RealType>
{
    public:
    template<typename... Args>
    ThrustDeviceWrapper(Args&&... args) :
    	thrust::device_vector<RealType>(std::forward<Args>(args)...){}

    const RealType* data() const
    {
        return thrust::raw_pointer_cast(
        	thrust::device_vector<RealType>::data());
    }
};
\end{verbatim}
Similarly, for Kokkos, a thin wrapper around \verb+Kokkos::vector<Real>+ can be used:
\begin{verbatim}
template <typename Real>
class KokkosVectorWrapper : public Kokkos::vector<Real>
{
    using KV = Kokkos::vector<Real>;

    public:

    template<typename... Args>
    KokkosVectorWrapper(Args&&... args) : KV(std::forward<Args>(args)...){}

    void pxr_sync()
    {
        Kokkos::deep_copy(KV::d_view, KV::h_view);
    }

    const Real* data() const
    {
        return KV::d_view.data();
    }
};
\end{verbatim}
PICSAR-QED calls the method \verb|pxr_sync()| whenever needed, if it detects that a type having such a method is used as the underlying vector type. \\
On CPUs, lookup tables can be passed directly as a constant reference to \emph{runtime} functions. On GPUs, an additional step is needed: each table can export a ``table view'', which is a lookup table of the same kind internally using non-owning pointers to the raw table data. This ``table view'' can be safely passed by copy to GPU kernels. This more complex procedure also works on CPUs. Therefore, if libraries such as AMReX or Kokkos are used, the same code can be compiled for both GPUs and CPUs.

\section*{Appendix C: technical details on PICSAR-QED benchmarks}
Table \ref{tab:comptab} reports several details (compiler versions, compilation options \ldots) on the configurations used for the benchmarks discussed in section \ref{sec:perftest}.
To ensure reproducibility of our results, we also provide the \emph{git commit number} corresponding to the version of the software that we have used in this work:
\begin{itemize}
\item PICSAR-QED~\cite{picsarqed_repo} : \verb|c16b642e3dcf860480dd1dd21cefa3874f395773|
\item WarpX~\cite{warpx_repo} : \verb|b83f2949a1ac2eed003e991e9653b8427716bf14| 
\item AMReX~\cite{amrex_repo} : \verb|b15b1cf8d282cbb2c0d0bc0c7b049a79375ea63c|
\end{itemize}
The code used for the performance benchmarks will be made available on the repository of PICSAR-QED.

\begin{table}
\begin{tabular}{|lp{4.2in}|}
\hline 
\multicolumn{2}{|l|}{\bf NVIDIA Quadro GV100: CUDA} \\
\textbf{\textit{GPU details}}: & {NVIDIA Quadro GV100, 32GB RAM, \newline Driver Version: 450.51.05, CUDA version: 11.0} \\
\textbf{\textit{Compilers}}: & {\verb-g++ (GCC) 7.3.1- (host), \verb-nvcc v11.0.194-}  \\
\textbf{\textit{Optimizations}}: & \verb+--gpu-architecture=sm_70+ \verb+-O3+ \verb#--use_fast_math# \\
\hline
\hline
\multicolumn{2}{|l|}{\bf NVIDIA Quadro GV100: Kokkos+CUDA} \\
\textbf{\textit{GPU details}}: & {(see above)} \\
\textbf{\textit{Compilers}}: & {\verb-g++ (GCC) 7.3.1- (host), \verb-nvcc v11.0.194-}  \\
\textbf{\textit{Kokkos version}}: & \verb+Kokkos tag: 3.3.01+ \\
\textbf{\textit{Optimizations}}: & {\verb+-O3+, \verb+Kokkos_ARCH_VOLTA70=ON+,
\verb+Kokkos_ENABLE_CUDA=ON+, \verb+Kokkos_ENABLE_CUDA_CONSTEXPR=ON+, 
\verb+Kokkos_ENABLE_CUDA_LAMBDA=ON+, \verb+Kokkos_ENABLE_AGGRESSIVE_VECTORIZATION+} \\
\hline
\hline
\multicolumn{2}{|l|}{\bf 2 x Intel Xeon Gold 6152: OpenMP (+precise)} \\
\textbf{\textit{Compilers}}: & Intel(R) oneAPI DPC++/C++ Compiler 2021.3.0 \\
\textbf{\textit{Optimizations}}: & {\verb+-O3 -fiopenmp -march=native -mtune=native+
\newline \verb+-fp-model=precise+} \\
\textbf{\textit{OpenMP options}}: & {\verb+OMP_PROC_BIND=spread+} \\
\hline
\hline
\multicolumn{2}{|l|}{2 x \bf Intel Xeon Gold 6152: Kokkos+OpenMP} \\
\textbf{\textit{Compilers}}: & \verb-g++ (Spack GCC) 11.1.0- (with \verb|Graphite|) \\
\textbf{\textit{Kokkos version}}: & \verb+Kokkos tag: 3.3.01+ \\
\textbf{\textit{Optimizations}}: & {
\verb+-O3+,
\verb+Kokkos_ARCH_SKX=ON+, 
\verb+Kokkos_ENABLE_OPENMP=ON+, 
\verb+Kokkos_ENABLE_AGGRESSIVE_VECTORIZATION+} \\
\textbf{\textit{OpenMP options}}: & {\verb+OMP_PROC_BIND=spread+} \\
\hline
\hline
\multicolumn{2}{|l|}{\bf AMD EPYC 7302 : OpenMP (with or without -fast-math)} \\
\textbf{\textit{Compilers}}: & \verb|AMD clang version 12.0.0|  \\
\textbf{\textit{Optimizations}}: & {\verb+-O3 (-Ofast -ffast-math) -fopenmp -march=znver3+} \\
\textbf{\textit{OpenMP options}}: & {\verb+OMP_PROC_BIND=spread+} \\
\textbf{\textit{Notes}}: &  {Lookup tables generation not compiled with fast-math.} \\
\hline
\hline
\multicolumn{2}{|l|}{\bf AMD EPYC 7302 : Kokkos+OpenMP} \\
\textbf{\textit{Compiler versions}}: & \verb|AMD clang version 12.0.0| \\
\textbf{\textit{Kokkos version}}: & \verb+Kokkos tag: 3.3.01+ \\
\textbf{\textit{Optimizations}}: & {\verb+-O3 -march=znver2 -mtune=znver2+, 
\verb+Kokkos_ENABLE_OPENMP=ON+},\\
&
{\verb+Kokkos_ENABLE_AGGRESSIVE_VECTORIZATION+} \\
\textbf{\textit{OpenMP options}}: & {\verb+OMP_PROC_BIND=spread+} \\
\hline
\end{tabular}
\caption{Details on the configurations used for the performance benchmarks. \label{tab:comptab}}
\end{table}

\section*{Acknowledgments}
This research used the open-source particle-in-cell code WarpX \url{https://github.com/ECP-WarpX/WarpX}, primarily funded by the US DOE Exascale Computing Project (17-SC-20-SC), a collaborative effort of the U.S. Department of Energy Office of Science and the National Nuclear Security Administration. We acknowledge all WarpX and PICSAR contributors. This research used resources of the Oak Ridge Leadership Computing Facility at the Oak Ridge National Laboratory, which is supported by the Office of Science of the U.S. Department of Energy under Contract No.DE-AC05-00OR22725. We acknowledge the support of the US DOE Exascale Computing Project and of the Director, Office of Science, Office of High Energy Physics, of the U.S. Department of Energy under Contract No. DEAC02-05CH11231. This work was supported by the French National Research Agency (ANR) T-ERC  program (Grant No. ANR-18-ERC2-0002). We also acknowledge the financial support of the Cross-Disciplinary Program on Numerical Simulation of CEA, the French Alternative Energies and Atomic Energy Commission. This project has received funding from the European Union’s Horizon 2020 research and innovation program under grant agreement No. 871072. The authors would like to thank Dr. Mathieu Lobet (Maison de la Simulation, CEA-Saclay, France) for useful discussions concerning the implementation of QED effects in PIC codes.

\printbibliography

\end{document}